\documentclass[prd,twocolumn,nofootinbib,showpacs]{revtex4}
\usepackage{epsfig}

\begin{document}

\title{The universe as a black hole in isotropic coordinates}
\author{Nikodem J. Pop{\l}awski}
\affiliation{Department of Physics, Indiana University,
727 East Third Street, Bloomington, IN 47405, USA}
\date{\today}

\begin{abstract}
We show that the radial geodesic motion of a particle inside a black hole in isotropic coordinates (the Einstein-Rosen bridge) is physically different from the radial motion inside a Schwarzschild black hole.
A particle enters the interior region of an Einstein-Rosen black hole which is regular and physically equivalent to the asymptotically flat exterior of a white hole, and the particle's proper time extends to infinity.
Because the motion across the Einstein-Rosen bridge is unidirectional, and the surface of a black hole is the event horizon for distant observers, an Einstein-Rosen black hole is indistinguishable from a Schwarzschild black hole for such observers.
Observers inside an Einstein-Rosen black hole perceive its interior as a closed universe that began when the black hole formed, with an initial radius equal to the Schwarzschild radius of the black hole $r_g$, and with an initial accelerated expansion.
Therefore the model of a universe as a black hole in isotropic coordinates explains the origin of cosmic inflation.
We show that this kind of inflation corresponds to the effective cosmological constant $\Lambda=3/r_g^2$, which, for the smallest astrophysical black holes, is $\sim10^{-8}\,\mbox{m}^{-2}$.
If we assume that our Universe is the interior of an Einstein-Rosen black hole, astronomical observations give the time of inflation $\sim10^{-3}\,\mbox{s}$ and the size of the Universe at the end of the inflationary epoch $\sim10^{32}\,\mbox{m}$.

\end{abstract}

\maketitle

The interval of the static, spherically symmetric, gravitational field in vacuum, expressed in isotropic coordinates, was found by Weyl \cite{Weyl}:
\begin{equation}
ds^2=\frac{(1-r_g/4r)^2}{(1+r_g/4r)^2}c^2 dt^2-(1+r_g/4r)^4(dr^2+r^2 d\Omega^2),
\label{interv1}
\end{equation}
where $0\le r<\infty$ is the radial coordinate, $d\Omega$ is the element of the solid angle, and $r_g=2GM/c^2$ is the Schwarzschild radius.
This metric does not change its form under the coordinate transformation:
\begin{equation}
r\rightarrow\frac{r_g^2}{16r},
\label{transf}
\end{equation}
and is galilean for $r\rightarrow\infty$.
Therefore it is also galilean for $r\rightarrow 0$, describing the Einstein-Rosen bridge between two, asymptotically flat, identical sheets, connected at the singular (det$\,g_{\mu\nu}=0$) surface $r=r_g/4$ \cite{ER,FW}.
The nonzero components of the Riemann curvature tensor for this metric are given by:
\begin{eqnarray}
& & R^{0\theta}_{\phantom{0\theta}0\theta}=R^{0\phi}_{\phantom{0\phi}0\phi}=R^{r\theta}_{\phantom{r\theta}r\theta}=R^{r\phi}_{\phantom{r\phi}r\phi}=\frac{r_g}{2r^3(1+r_g/4r)^6}, \nonumber \\
& & R^{0r}_{\phantom{0r}0r}=R^{\theta\phi}_{\phantom{\theta\phi}\theta\phi}=-\frac{r_g}{r^3(1+r_g/4r)^6},
\label{curv}
\end{eqnarray}
so the Kretschmann scalar is nonsingular everywhere: $R_{\mu\nu\rho\sigma}R^{\mu\nu\rho\sigma}=12r_g^2 r^{-6}(1+r_g/4r)^{-12}$, going to zero as $r\rightarrow\infty$ and $r\rightarrow 0$.
The Einstein-Rosen metric for $r>r_g/4$ describes the exterior of a Schwarzschild black hole (the transformation of the radial coordinate $r(1+r_g/4r)^2\rightarrow r$ brings the interval (\ref{interv1}) into the standard Schwarzschild form \cite{Schw}).
The spacetime given by the metric (\ref{interv1}) for $r<r_g/4$ is regarded by an observer at $r>r_g/4$ as the interior of a black hole, although it really is, as we will see below, the (spherical) mirror image of the other exterior sheet.
This situation is analogous to the method of images in electrostatics, where the interaction between an electric charge situated at a distance $r$ from the center of a conducting sphere of radius $R<r$ is equivalent to the interaction of the same charge with a charge of the opposite sign situated inside this sphere at a distance $R^2/r$ from its center.
The radius $R$ corresponds to the Schwarzschild surface in isotropic coordinates, $r=r_g/4$.

Consider a massive particle moving radially in the gravitational field described by the metric (\ref{interv1}).
For brevity, we use:
\begin{equation}
h=g_{00}=\frac{(1-r_g/4r)^2}{(1+r_g/4r)^2},\,\,\,\,f=-g_{rr}=(1+r_g/4r)^4.
\label{not}
\end{equation}
The motion of the particle is given by the radial geodesic equations.
If the particle is at rest at $r=r_0$, then these equations are:
\begin{eqnarray}
& & \frac{dt}{d\tau}=u^0=\sqrt{h_0}/h, \label{mot1} \\
& & \frac{dr}{cd\tau}=u^r=\epsilon(h_0 h^{-1} f^{-1}-f^{-1})^{1/2},
\label{mot2}
\end{eqnarray}
where $\tau$ is the proper time of the particle, $h_0=h|_{r=r_0}$, and $\epsilon=-1\,(+1)$ for an infalling (outgoing) motion.
Consider a particle falling into a black hole, $\epsilon=-1$.
As $r\rightarrow r_g/4$, $h$ goes to zero and $f\rightarrow 16$, so both $u^0$ and $u^r$ become infinite.
Even if the initial motion were not purely radial, the components $u^0,\,u^r$ would still become infinite at $r=r_g/4$, with $u^\theta,\,u^\phi$ remaining finite.
Therefore, each motion of a massive particle becomes effectively radial at the surface $r=r_g/4$.

A distant observer situated in a nearly galilean spacetime measures the velocity of the particle as:
\begin{equation}
v_d=\frac{dr}{dt}=c\frac{u^r}{u^0}=-c\frac{(h_0 f^{-1}h-f^{-1}h^2)^{1/2}}{\sqrt{h_0}}.
\label{mot3}
\end{equation}
As $r\rightarrow r_g/4$, $v_d$ goes to zero.
Writing $r=r_g/4+\xi$, where $0<\xi\ll r_g$, gives $v_d\approx -c\xi/2r_g$ and thus $r-r_g/4\sim\mbox{exp}(-ct/2r_g)$, so the particle reaches the surface of a black hole $r=r_g/4$ after an infinite time $t$.
This surface is an event horizon for a distant observer, as it is for the standard Schwarzschild metric \cite{Lem}.
The proper time $\Delta\tau$ of the particle for moving radially from $r=r_0$ to $r=r_g/4$ is finite, which can be shown by considering $r_0=r_g/4+\xi$:
\begin{equation}
c\Delta\tau=\int_{r_g/4}^{r_g/4+\xi}dr/u^r\approx r_g.
\label{mot4}
\end{equation}
After reaching the surface $r=r_g/4$, the particle continues moving; its radial coordinate $r$ decreases to $r_1=r_g^2/16r_0\approx r_g/4-\xi$ (at which $u^r=0$) in a proper time $\Delta\tau=r_g/c$.
The radial motion of a massive particle (in terms of the proper time) in the spacetime (\ref{interv1}) for $r\le r_g/4$ is the (spherical) mirror image of the particle's motion for $r\ge r_g/4$.
Applying the transformation (\ref{transf}) to the motion for $r\le r_g/4$ (inside a black hole) shows that this motion is equivalent to the outgoing motion, in terms of the new radial coordinate $r'=r_g^2/16r$, from a white hole (the time reversal of a black hole).

The local velocity of the particle $v_l$, measured in terms of the proper time, as determined by static clocks synchronized along the trajectory of the particle, is related to $u^0$ by $u^0=(h(1-v_l^2/c^2))^{-1/2}$ \cite{LL}.
As the particle moves from $r=r_0$ to $r=r_g/4$, $v_l$ increases from zero to $c$, and as the particle moves from $r=r_g/4$ to $r=r_1$, $v_l$ decreases to zero.
In a Schwarzschild field, $v_l$ exceeds $c$ inside a black hole, which does not violate Einstein's theory of relativity because the interior of a Schwarzschild black hole is not static and neither can be clocks synchronized along the trajectory of the particle.

A maximal extension of the Schwarzschild metric \cite{KS} shows that a massive particle cannot travel in this spacetime from (exterior) region I to (exterior) region III without violating causality: the Schwarzschild bridge is not traversable \cite{FW,MT}.
Such a particle either remains in region I or moves to (interior) region II, where it reaches the central singularity.
Unlike inside a Schwarzschild black hole, where the proper time ends at the central singularity, the proper time of a particle moving into an Einstein-Rosen black hole (bridge) does not end (the particle's geodesic is complete).
Such a particle enters the interior of an Einstein-Rosen black hole (with no possibility of coming back to the exterior sheet), which has no curvature singularities and is mathematically equivalent to the asymptotically flat exterior of a white hole \cite{NN}.
Therefore an Einstein-Rosen black hole is physically different from a Schwarzschild black hole with regard to the nature of the interior sheet, and equivalent to it with regard to the nature of the exterior sheet.
For distant observers, that can only see the exterior sheet, the two black holes are indistinguishable.
Another difference comes from the integral of the time-time component of the gravitational energy-momentum psedudotensor (either Einstein or Landau-Lifshitz) over the interior hypersurface, that is, the total energy of the system.
This energy equals the sensible physical value $r_gc^4/2G=Mc^2$ for an Einstein-Rosen black hole, but diverges for a Schwarzschild black hole.

We can regard the interior of an Einstein-Rosen black hole as a new (closed) universe that began with the formation of the black hole from a supernova explosion, in a globular cluster or at the center of a galaxy.
This interpretation \cite{wh} is suggested by the gravitational collapse of a homogeneous sphere of dust in Tolman's coordinates \cite{OS}, which has a solution of form of the Friedmann metric for a closed isotropic universe \cite{LLF}.
The motion of the boundary of such a sphere is physically equivalent to the radial motion of a particle in the Schwarzschild spacetime, ending at the central singularity \cite{LLF}.
In order to analyze the gravitational collapse of a homogeneous dust sphere with the boundary moving like a particle in the Einstein-Rosen spacetime (such a sphere is a different solution of the Einstein equations than Tolman's sphere in the same sense that the Schwarzschild metric is different from the Einstein-Rosen metric), one can generalize Wyman's interior solution for a sphere of perfect fluid in isotropic coordinates (which is equivalent to the interior solution in standard Schwarzschild coordinates \cite{Tol}) to nonstatic cases \cite{isot}.
The vanishing of det$\,g_{\mu\nu}$ at the surface $r=r_g/4$ is not a physical singularity, it only means that one can introduce a locally inertial frame of reference everywhere except at the boundary between two universes.
The nature of the interior of a black hole \cite{EF} has not been satisfactorily determined and is open to considerable debate \cite{int}.

\begin{figure}[ht]
\centering
\includegraphics[width=3in]{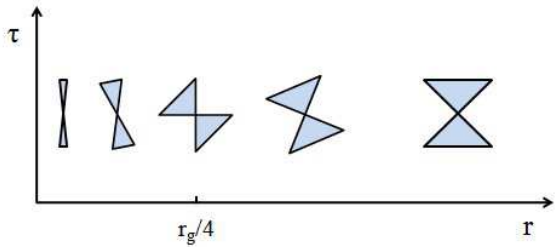}
\caption{Light cones in the gravitational field represented by the metric (\ref{interv2}).}
\label{fig}
\end{figure}

Consider radially moving geodesic clocks that are falling into a black hole from the isotropic radius $r_0$, and synchronized such that the time of each clock at the instant of release equals the proper time $\tau_0$ of a clock at rest that remains fixed at $r_0$.
The time at any event is taken to be equal to the proper time on the radially falling clock that is coincident with this event, as in \cite{GH}.
Equations (\ref{mot1}) and (\ref{mot2}) give (for $\epsilon=-1$):
\begin{equation}
cd\tau=\sqrt{h_0}cdt+\sqrt{f(h_0-h)/h}dr,
\label{mot5}
\end{equation}
which, using $\tau_0=\sqrt{h_0}t_0$, integrates to:
\begin{equation}
c\tau=\sqrt{h_0}ct+\int_{r_0}^{r}\sqrt{f(h_0-h)/h}dr,
\label{mot6}
\end{equation}
giving the transformation between the coordinates $(r,t)$ and $(r,\tau)$.
In terms of $\tau$, the metric (\ref{interv1}) becomes:
\begin{equation}
ds^2=h(cd\tau-\sqrt{f(h_0-h)/h}dr)^2/h_0-fdr^2-fr^2d\Omega^2.
\label{interv2}
\end{equation}
Radial null geodesics are given by $ds=0$ and $d\Omega=0$:
\begin{equation}
\frac{cd\tau}{dr}=\sqrt{fh_0/h}(\sqrt{1-h/h_0}\pm1).
\label{nul}
\end{equation}
The plus (minus) sign corresponds to an outgoing (infalling) null geodesic.
For $r\rightarrow\infty$, the spacetime is galilean and $cd\tau/dr=\pm1$.
For $r=r_g/4$, $cd\tau/dr=\infty$ for the outgoing null geodesic and $0$ for the infalling one, as shown in Figure \ref{fig}.
For $r=0$, $cd\tau/dr=\pm\infty$; however, in terms of the new radial coordinate $r'=r_g^2/16r$ (\ref{transf}), we obtain $cd\tau/dr'=\pm1$ (galilean spacetime).
Massive particles can move in both radial directions, except at the unidirectional surface $r=r_g/4$, where only infalling geodesics (decreasing $r$) lie inside the light cone.

The metric (\ref{interv2}) can be written as:
\begin{equation}
ds^2=c^2d\tau^2-(\sqrt{h_0-h}cd\tau+\sqrt{fh}dr)^2/h_0-fr^2d\Omega^2.
\label{interv3}
\end{equation}
Taking $h_0=1$, which corresponds to $r_0\rightarrow\infty$, reduces Eq. (\ref{interv3}) into:
\begin{equation}
ds^2=c^2d\tau^2-(\sqrt{1-h}cd\tau+\sqrt{fh}dr)^2-fr^2d\Omega^2.
\label{interv4}
\end{equation}
Define a new radial coordinate $R(\tau,r)$ such that:
\begin{equation}
dR=\sqrt{1-h}cd\tau+\sqrt{fh}dr.
\label{rad1}
\end{equation}
The interior of an Einstein-Rosen black hole will be an isotropic universe if also:
\begin{equation}
R^2=fr^2.
\label{rad2}
\end{equation}
Equations (\ref{rad1}) and (\ref{rad2}) give:
\begin{equation}
d(\mbox{ln}R)=\frac{|\sqrt{1-h}cd\tau+\sqrt{fh}dr|}{\sqrt{f}r}.
\label{rad3}
\end{equation}
If $r=r_g/4$ then $f=16$ and $h=0$, so Eq. (\ref{rad3}) gives:
\begin{equation}
R\sim e^{c\tau/r_g},
\label{sol1}
\end{equation}
which represents an accelerated expansion of a de Sitter universe ($R\sim\mbox{exp}(\sqrt{\Lambda/3}c\tau)$) with the cosmological constant (the motion of the boundary of a collapsing sphere is dynamically equivalent to the radial motion of a particle in the gravitational field of this sphere):
\begin{equation}
\Lambda=\frac{3}{r_g^2}.
\label{sol2}
\end{equation}
If $r=0$ then $h=1$, so Eq. (\ref{rad3}) gives:
\begin{equation}
R\sim r^{-1},
\label{sol3}
\end{equation}
which is consistent with interpreting $1/r$ as the physical radial coordinate in the interior sheet (the metric (\ref{interv1}) is asymptotically flat in terms of the radial coordinate $r_g^2/16r$).

If the Universe is the interior of an Einstein-Rosen black hole and began with the formation of the black hole from a supernova explosion in the center of a galaxy, then $r_g$ is on the order of the size of a star that is massive enough to form a black hole, that is, $r_g\sim 10\,\mbox{km}$.
Consequently, the initial value of the cosmological constant $\Lambda$ is on the order of $10^{-8}\,\mbox{m}^{-2}$.
Such a large cosmological constant may be associated with cosmic inflation in the early Universe.
Therefore the model of a universe as a black hole in isotropic coordinates explains the origin of inflation.
Because astronomical observations suggest that the Universe in the inflationary epoch expanded $28$ or more orders of magnitude \cite{GL}, inflation caused by this Schwarzschild-related cosmological constant took at least $\tau_i=28\mbox{ln}10r_g/c\sim10^{-3}\,\mbox{s}$.
The size of the whole Universe at the end of the inflationary epoch was then larger than $10^{32}\,\mbox{m}$.

If the Einstein-Rosen black hole that formed our Universe were alone, then the value of the cosmological constant would have remained large and no structures would have formed.
However, if this black hole were densely surrounded by stars, the value of $r_g$ would have been increasing as stellar matter falls into the black hole, and $\Lambda$ would have been decreasing.
In order to find the final value of $\Lambda$ after the black hole has consumed all of available matter, one should generalize the above considerations of the Universe as an Einstein-Rosen white hole to the presence of matter.
Without such a generalization, Eq. (\ref{sol2}) would yield, with the current value $\Lambda=1.3\times10^{-52}\,\mbox{m}^{-2}$ \cite{cosm}, the current value of the radius of the Einstein-Rosen white hole that is our Universe: $r_g=1.5\times10^{26}\,\mbox{m}$.
This value of $r_g$ is on the order of the size of the observable Universe, which is at least $r_g=3.7\times10^{26}\,\mbox{m}$ \cite{rad}, and $\sim10^6$ times smaller than the whole Universe at the end of the inflationary epoch.
Therefore, in the presence of matter, we expect Eq. (\ref{sol2}) to be modified such that the current value of $\Lambda$ resulting from this equation will be $\sim10^6$ times smaller.

\end{document}